\documentclass[11pt]{article}
\usepackage{amsfonts,amsthm,amssymb}
\usepackage{cite}
\usepackage[pdftex]{graphicx}
\usepackage{color}
\usepackage[fleqn]{amsmath}
\usepackage{enumitem}
\usepackage[leftcaption]{sidecap}
\usepackage{hyperref}
\newcommand{\be}{\begin{eqnarray}}
\newcommand{\ee}{\end{eqnarray}}
\newcommand{\bez}{\begin{eqnarray*}}
\newcommand{\eez}{\end{eqnarray*}}

\allowdisplaybreaks

\begin{document}

\title{Obituary: Aristophanes Dimakis} 
\author{
 \sc Folkert M\"uller-Hoissen \\ \small 
 Institut für Theoretische Physik, Friedrich-Hund-Platz 1,
 37077 G\"ottingen, Germany \\
 \small  folkert.mueller-hoissen@theorie.physik.uni-goettingen.de
}

\date{ }

\maketitle

\begin{abstract}
The theoretical physicist and mathematician Aristophanes Dimakis passed away on July 8, 2021, at the age of 68, in Athens, Greece. We briefly review his life, career and scientific achievements.
\end{abstract}    

\noindent
We mourn the loss of a very close friend and esteemed colleague. 

\begin{figure}[ht]
\begin{center}
\includegraphics[scale=.09]{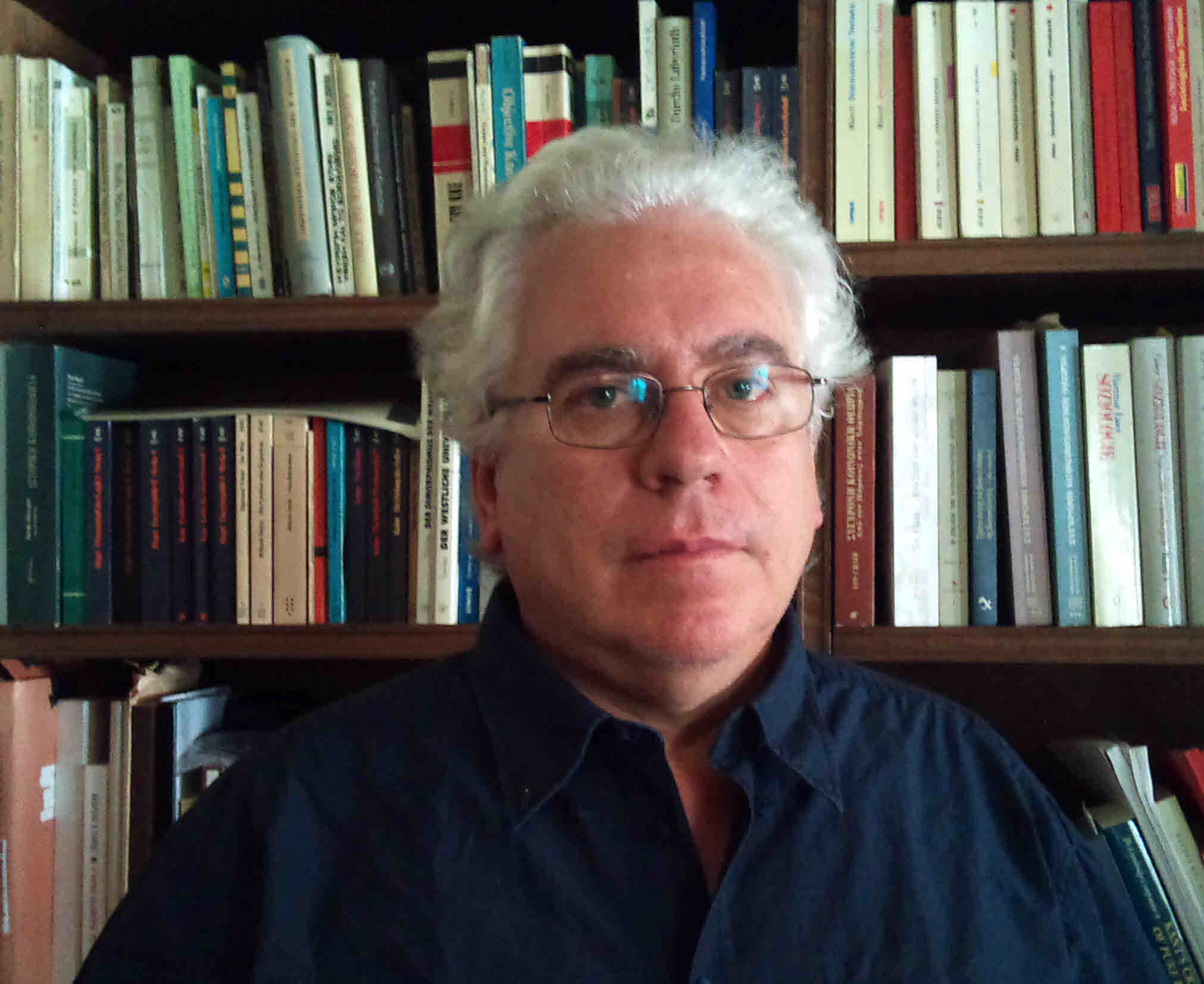}
\caption{Aristophanes Dimakis in 2011 in his apartment in Athens.}
\end{center}
\end{figure}

\section{Life and career of Aristophanes Dimakis}

Aristophanes was born in Orestiada, Greece, on March 24, 1953. He attended elementary school in Orestiada and later in Kavala, then entered high school in Kavala, later in Athens, where he graduated from it in 1971. 
In autumn 1971 he passed the entrance examination of the National and Kapodistrian University of Athens 
and began to study physics. 

February 1973 saw the rise of actions by students in Athens against the military dictatorship in those days, 
notably the occupation of the building of the Law School, and Aristophanes took part in it. 
All this escalated in the ``Polytechnic uprising". On November 14, 1973, hundreds of students 
occupied the National Technical University of Athens (Polytechnion),  barricaded 
themselves in it and even built a radio station. The rebellion ended bloodily in the early morning 
of November 17, 1973, when a tank drove over the gate of the Polytechnic  and snipers shot 
from the roofs of nearby buildings. Unlike many others, Aristophanes survived, was not even injured 
and managed to escape arrest. November 17 is still a commemoration day in Greece.

At the end of 1976 Aristophanes received a diploma in physics from the University of Athens. 
In spring 1977 he went to Göttingen, Germany, where he wanted to go ahead for receiving a PhD degree. But a Greek diploma was rarely accepted in those days, so first he had to write another diploma thesis, for which he chose Professor Hubert Goenner as his supervisor, who headed a group at the Institute for Theoretical Physics, concentrating on topics in General Relativity and theories of gravitation. At this time, I also worked on a diploma thesis in the same group. We shared a room in the Institute for Theoretical Physics, in those days located in the Bunsenstrasse, and became friends and lifelong collaborators. We both received our PhD degrees in 1983. Aristophanes' German diploma thesis was entitled ``Newtonsche und post-Newtonsche Näherung im Rahmen einer Gravitationstheorie ohne minimale Kopplung" (Newtonian and post-Newtonian approximation in the framework of a theory of gravity without minimal coupling). His doctoral thesis had the title  ``Geometrische Behandlung von Clifford-Algebren und Spinoren mit Anwendungen auf die Dynamik von Spinteilchen" (Geometric treatment of Clifford algebras and spinors with applications to the dynamics of particles with spin). From April 1978 to July 1980 he also worked as a tutor at the Institute for Theoretical Physics. In November 1981 he became a scientific employee, financed via a project funded by the Deutsche Forschungsgemeinschaft (DFG) and headed by Hubert Goenner.

So far Aristophanes had managed to postpone his military service in Greece. Finally he was able to reduce it to a few 
months which he completed at the end of 1987. He arranged this between some non-permanent positions at the Institute for Theoretical Physics in Göttingen. On February 16, 1990, Aristophanes took up a position at the Department of Mathematics at the University of Crete in Iraklion. In those days he has been particularly supported by Basilis Xanthopoulos, who was assassinated on November 27, 1990, during a seminar, which Aristophanes also attended. Yet another colleague was killed in this rampage of a student and Aristophanes has been lucky to survive this. I had visited him a year ago on Crete and remember that during a seminar delivered by Vladimir E. Zakharov we heared a loud shot. Just for fun I said  ``Oh, someone has been shot!" and learned later that guns are indeed not at all rare on this island, quite well-known for its vendettas.  

On September 1, 1996, he joined the Aegean University on Samos and on September 25, 2000,  he moved to the Department of Financial and Management Engineering on Chios, where he taught mathematics to future engineers and continued his research in mathematical physics. Almost at the age of fifty, on August 23, 2002, he finally received a permanent position as a full professor.  He also served for many years as head of the department and retired  in September 2020.

Aristophanes' research started in General Relativity. Later he turned to Noncommutative Geometry and finally concentrated on mathematical aspects of Integrable Systems. The variety of ideas he developed in his publications is quite amazing. 

Aristophanes died on July 8, 2021, in a hospital in Athens. In April he had been diagnosed with lung cancer that progressed rapidly. His appointment as ``emeritus" had taken a while and only reached him when he was already under intensive care in the hospital. 

Aristophanes was a very gentle person. Though not quite a sociable person (at least outside of Greece and his familiar environment of close friends), those who met him very much appreciated his good sense of humour and broad knowledge reaching far beyond the realm of science. He loved to tackle mathematical problems and often found an ingenious solution within a surprisingly short time, during which he worked with enormous concentration. Surely mathematics made up the main part of his life. He also had the attitude to quickly turn to a new topic, once he had revealed the secrets of the former. In this way, many of his results got somehow lost in his folders. In March 2021 he wrote to me  ``I feel empty of research ideas. This is not healthy." He was steadily looking for new mathematical problems to tackle. Certainly we would have seen many more deep results had he not gone so early.

He left behind his spouse Aliki Lavranou, since 2020 professor at the School of Social Sciences of Panteion University in Athens. They had met in Göttingen in the early 80th and spent about 40 years together. Later they shared an apartment in Athens and for a long time both had to fly weekly back and forth between Athens and different Greek islands for delivering their teaching. I have to thank Aliki for providing me with some data about Aristophanes' life.

\section{A r\'esum\'e of his scientific work}
After a short mathematical excursion to economics \cite{DS80a,D80b}, Aristophanes addressed problems in General Relativity and its extensions with \emph{torsion}, the simplest being Einstein-Cartan theory. Whereas classical particles cannot detect torsion, Dirac (spinor) fields, which model fermions like electrons and neutrinos,  in principle can. In those days one studied the (speculative) possibility of classical Dirac fields as sources in Einstein's equations. A special case occurs if a Dirac field does \emph{not} contribute to the energy-momentum tensor, the source of the gravitational field. These ``ghost" Dirac fields were studied in \cite{DMH82,DMH83}. In \cite{DMH85}, the Einstein-Cartan equation with vanishing energy-momentum tensor and a massive Dirac field was shown to reduce to covariant equations on the group $SL(2,\mathbb{R})$, a nice application of differential geometry on Lie groups.

Newton's theory of gravity can be cast into a differential geometric form analogously to General Relativity. Between this \emph{Newton-Cartan theory} and General Relativity exists an infinity of higher order approximations, which also have a geometric formulation. The publication \cite{D85} was mainly based on Aristophanes' diploma thesis. 

During his PhD work, Aristophanes became a top expert on \emph{Clifford algebras}. His first publications \cite{D86,D89c} in this area were inspired by work of David Hestenes. In \cite{DMH91} a comprehensive formalism based on Clifford algebras was developed, in particular (but not only) to  ease complicated computations in the area of higher-dimensional gravity theories. The term ``clifform'' was coined to abbreviate the notion ``Clifford-algebra-valued differential form''.

In \cite{D89a,D89b} studies of the initial value problem for theories of gravity with torsion was presented.

The work \cite{DMH89} has been motivated by Edward Witten's famous spinor-based proof of the \emph{positive energy theorem} in General Relativity. A condition considered by James M. Nester to select a convenient  orthonormal frame field on a three-dimensional Riemannian manifold was shown to be equivalent to the (linear) Dirac equation. In \cite{DMH90} a proof of the positive energy theorem was presented using orthonormal frame fields instead of spinors.

Though Aristophanes and I did not publish any work about quantum gravity, the problem of unifying General Relativity with Quantum Mechanics very much occupied our thoughts in those days. When ``Noncommutative Geometry''  came up, we regarded it as a promising framework for such a unification. But soon we realized that, before addressing such a goal, we had to get a deeper understanding of this essentially new mathematics. We decided to consider simpler applications. 

Differential geometry underlies General Relativity, as well as the gauge theories that form the crucial structure of electrodynamics and elementary particle physics. The most basic structure is a manifold, or better the commutative associative algebra of (smooth) functions on it. Noncommutative Geometry generalizes it to a, in general, noncommutative  (mostly still associative) algebra. Required in addition is an analog $\Omega$ of the algebra of differential forms and, in particular, an analogue of the exterior derivative $d$, acting on $\Omega$ such that the (graded) Leibniz rule still holds. Further geometric structures, like analogues of connections (gauge fields), symplectic structures, notions of distance and metrics, can then be introduced on $\Omega$. 

In \cite{DMH92a,DMH93a} it has been explored in which sense Quantum Mechanics can be understood as ``noncommutative symplectic geometry". When we then realized that commutative algebras can carry non-standard differential calculi, specified by non-trivial commutation relations between elements of the algebra and their  differentials, this opened up a rich new world. For example, imposing
\be
             [x,dx] = \ell \, dx    \qquad \quad \ell \in \mathbb{R} \, , \quad \ell > 0 \, ,    \label{latt_calc}
\ee
on a commutative algebra generated by $x$,  this generalizes to $f(x) \, dx = dx \, f(x+\ell)$  for any function $f$ of $x$. By using the Leibniz rule, one then easily deduces that
\bez
           df = dx \, {1\over \ell} \,  \lbrack \, f(x+\ell) - f(x) \, \rbrack  
                =  {1\over \ell} \,  \lbrack \, f(x) - f(x-\ell) \, \rbrack \; dx \, ,
\eez
where left and right discrete derivatives show up. So this differential caluclus can be though of as living on a one-dimensional lattice with spacing $\ell$. This calculus and its higher-dimensional version have been explored in \cite{DMH92c,DMHS93a,DMHS93b}. More generally, differential calculi on finite sets, and geometric structures built on them, have been the subject of quite a number of further publications \cite{DMH93d,DMH94a,DMH94c,DMH94d,DMH95a,DMH95b,BDMHS96}, apart from those by other authors. In particular, we made efforts to develop ``discrete Riemannian geometry" in as close as possible analogy with differential geometry \cite{DMH99b,DMH03a,DMH03b}.

Relations of the form $[x^i , dx^j] = C^{ij}_k \, dx^k$ with constants $C^{ij}_k$, $i,j,k=1,\ldots,n$, were classified in \cite{DMHB95} up to $n=3$. An example from this class, different from (\ref{latt_calc}), turned out to describe stochastic processes on manifolds \cite{DMH93c,DMH94b}. A framework for kinetic theory of open systems, based on this ``noncommutative calculus'' was developed in \cite{DT96,D97,DT00}. In \cite{DMH92b} (also see \cite{DMH93b}) we related this calculus to physical theories postulating two notions of  time. We sent the manuscript to Nuclear Physics B, where is was delayed for about a year until we received a negative decision from an editor (without a report, but along with a negative general statement about Noncommutative Geometry). Anyway, we were proud of this work, but did not submit it again to a journal.

\begin{figure}
\begin{center}
\includegraphics[scale=.7]{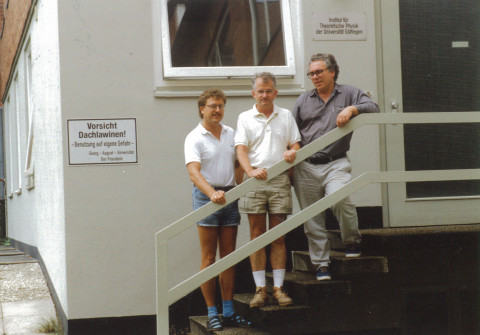}
\caption{Aristophanes Dimakis together with John Madore (middle) and Folkert M\"uller-Hoissen in 1995 in front of the Institute for Theoretical Physics in G\"ottingen, in those days located in the Bunsenstrasse. John, who also worked on General Relativity before he turned to Noncommutative Geometry,  passed away in August 2020.}
\end{center}
\end{figure}

Many completely integrable PDEs admit a differential-geometric zero curvature formulation, which suggested to explore a noncommutative version of the latter. This led us again into a rich new world. Besides some first explorations \cite{DMH96b}, we came across \emph{chiral models},
\bez
           d \star (g^{-1} \, dg)  = 0 \, , 
\eez
 where $g$ is an invertible matrix of functions on a 2-dimensional Riemannian space and $\star$ the Hodge star operator. Such models were known to possess an infinite number of conserved currents, which is an integrability feature. The goal was to generalize them to a framework of Noncommutative Geometry, by generalizing the underlying differential calculus and the star operator, while preserving integrability.  This was achieved in \cite{DMH96a,DMH97a,DMH97b,DMH98b}. For example, by deforming the standard differential calculus in one of the two dimensions to the lattice calculus (\ref{latt_calc}) and using a star operator that corresponds to that of 2-dimensional Minkowski space, one obtains the famous Toda lattice, which thus turned out to be a chiral model with respect to a noncommutative geometry.

A variety of studies followed, on Noncommutative Geometry \cite{DM96,D96,DMH04e}, discretization \`a la \emph{Umbral calculus} \cite{DMHS96}, Connes' distance function on discrete sets \cite{DMH98a}, soliton equations on a ``noncommutative space'' \cite{DMH00g,DMH00h,DMH01c}, \emph{Moyal deformation} of integrable models \cite{DMH00i,DMH01a,DMH04c,DMH04d,DMH04f,DMH05a,DMH07c}, automorphisms of real four-dimensional Lie algebras and characterization of four-dimensional homogeneous spaces \cite{DCP03}, ``functional representations" (generating equations) of integrable hierarchies \cite{DMH06b}, studies of KP and related hierarchies \cite{DMH07a,DMH07b,DMH09c,DMH09d,DMH08b,DMH08c}, relations between different hierarchies via deformation of multiplication \cite{DMH06c}. 

An excursion into \emph{nonassociativity} was made in \cite{DMH06a,DMH06d,DMH08a,DMH09a,DMH10a}. Let $f$ freely generate a nonassociative algebra $\mathbb{A}$.
Imposing the (``weak nonassociativity") condition $a((bc)d)-(a(bc))d=0$ for all $a,b,c,d \in \mathbb{A}$, the algebra admits an infinite sequence of derivations $\delta_n$, $n \in \mathbb{N}$, given by monomials in $f$. These derivations satisfy identities which are, via $\delta_n \mapsto \partial/\partial t_n$, in correspondence with equations of the KP hierarchy, with dependent variable in a noncommutative associative algebra. 
The associative subalgebra of $\mathbb{A}$ is isomorphic to the algebra of \emph{quasi-symmetric functions} and the latter carries the structure of an ``infinitesimal bialgebra'' \cite{DMH10a}. I'm not sure that the beauty of these results 
reached the world around us ...

In the case of the abovementioned chiral models, the integrability feature was the existence of an infinite tower of conservation laws. Integrable models (notably soliton equations) typically possess powerful solution-generating methods, which we considered as more useful. They can often be derived from a parameter-dependent zero curvature (Zakharov-Shabat) condition. In the framework of Noncommutative Geometry, we therefore started to explore the case of a linear dependence on such a parameter, which led to a structure which we called ``bidifferential calculus'' and which can be thought of as a generalization of \emph{Fr\"olicher-Nijenhuis theory} from Differential to Noncommutative Geometry. It turned out that indeed many integrable models (though probably not all) can in fact be treated in this framework 
\cite{DMH00b,DMH00c,DMH00d,DMH00e,DMH00f,DMH01a,DMH01b,DMH02,DMH09b,DMH10b,DMH10c,DKMH11,DMH13,DCMH16,DMH20a}.
A crucial point was that integrability features could be formulated in an extremely general way, only using simple calculation rules of bidifferential calculus. Treating a concrete model, i.e., an integrable differential or difference equation, meant to find a suitable realization of the graded algebra and two linear operators acting on it with very simple properties. It took a while, and several less satisfactory publications, until we came up in \cite{DMH13} with a sufficiently general ``abstract version'' of what is known as a ``binary Darboux transformation", a powerful solution-generating method for integrable models. 

\begin{figure}[ht]
\begin{center}
  \begin{minipage}[t]{6cm}
    \vspace*{-7cm}
    \includegraphics[scale=.42]{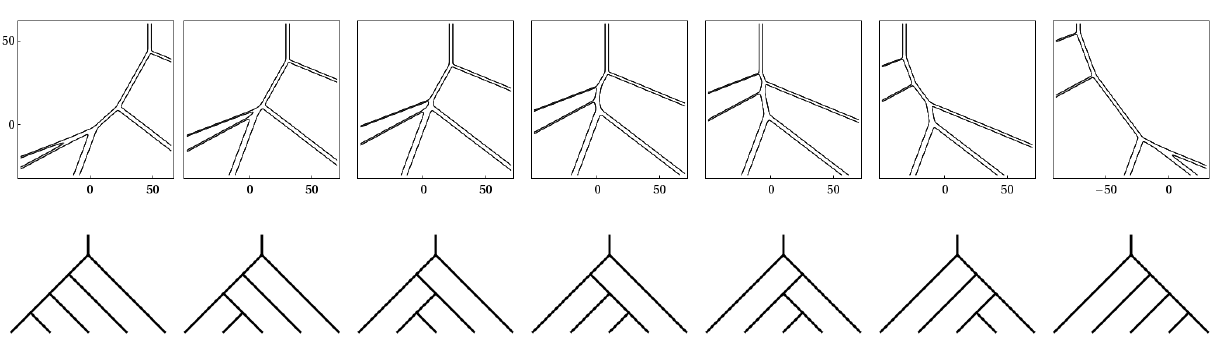}
  \end{minipage}
  \hspace{3cm}
 \begin{minipage}[t]{150pt}
   \includegraphics[scale=.5]{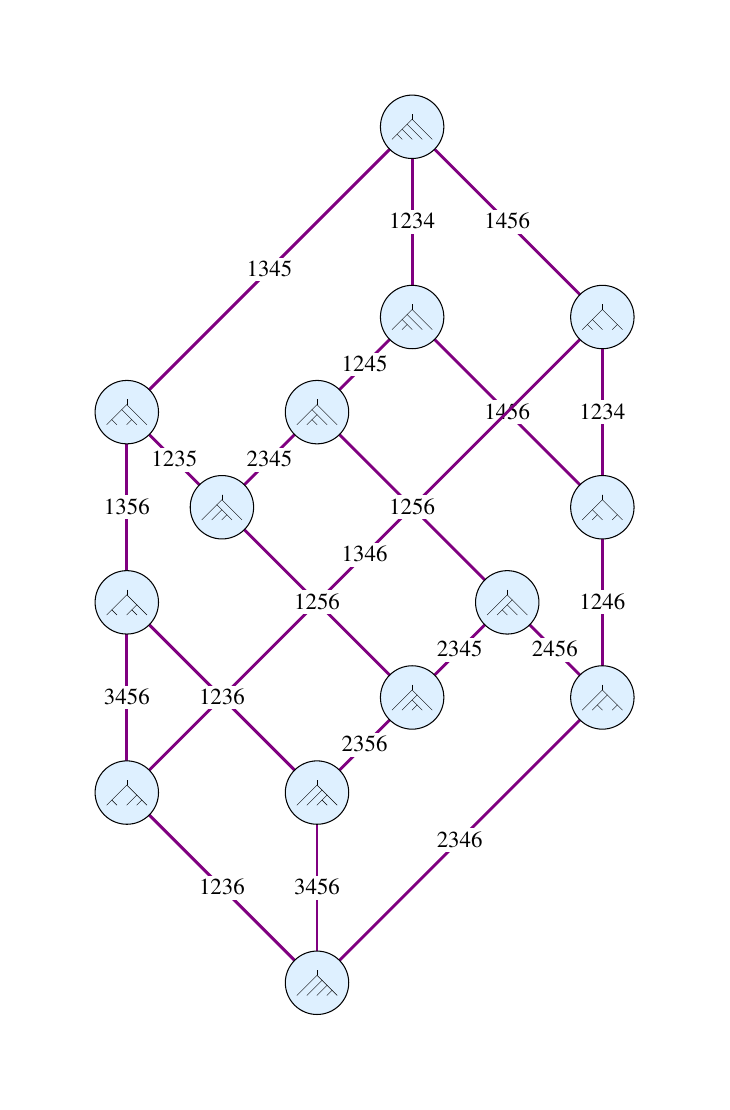}
  \end{minipage}
\caption{A sequence of contour plots of a soliton solution of the KP equation at consecutive values of time. The second row displays more clearly the corresponding rooted binary trees. The KP equation describes fluid surface wave, but it is difficult to generate such waves in order to observe the tree rotation in nature. The right figure displays a Tamari lattice which contains the path shown in the left figure. Regions in the plane separated by legs of a rooted binary tree are consecutively numbered from left to right counterclockwise. A sequence of numbers shown in the Tamari lattice indicates which of these regions are involved in the respective rooted binary tree transition via a right rotation. \label{fig:Tamari} }
\end{center}
\end{figure}

Solitons of integrable equations like Korteweg-deVries (KdV) or the Nonlinear Schr\"odinger (NLS) equation are spatially localized wave solutions in two space-time dimensions. Taking the wave crest limit of a 2-soliton solution leads to a graph with two incoming and two outgoing lines and an interaction vertex. More generally, such a ``particle picture" can be associated with solitons of certain integrable equations via a ``tropical limit'', so named because of relations with tropical mathematics. This also works for solitons of the KP equation, which lives in three space-time dimensions. It possesses a subclass of soliton solutions for which, at a fixed time, the tropical limit graph has the form of a rooted binary tree. Surprisingly, the time evolution then corresponds to a sequence of what is known as a ``right rotation" (of a ``leg", i.e., an edge) at some vertex. The concrete sequence of right rotations depends on parameters of the soliton solution. The various rooted binary trees with a fixed number of legs (corresponding to a fixed number of solitons) connected by single right rotations form a so-called \emph{Tamari lattice} (see Fig.~\ref{fig:Tamari}), a combinatorial structure expressing \emph{associativity} relations. All Tamari lattices are thus realized by soliton solutions of the KP equation \cite{DMH11} (the latter work has been chosen for IOP Select). There are also relations of this kind between so-called \emph{higher Tamari orders} and the KP hierarchy \cite{DMH12}. 

A follow-up work was the study \cite{DMH15a,DMH15b} of \emph{simplex and polygon equations}. Similar to the known relation between simplex equations (which include the Yang-Baxter equation) and \emph{higher Bruhat orders}, there is a corresponding relation between polygon equations (which include the pentagon equation) and \emph{higher Tamari orders}.
Aristophanes' contributions were the most crucial in these works. Moreover, he was a grandmaster of {\sc Mathematica} and used it in particular to generate a lot of beautiful illustrations for these publications. Aristophanes' last publication \cite{DK21}, jointly with Igor Korepanov, presented a large class of solutions of simplex and polygon equations with maps acting between direct sums of vector spaces.

\begin{figure}
\begin{center}
\begin{minipage}[b]{1cm}
\bez
  \begin{array}{cc} y & \\ \uparrow & \\ & \rightarrow x \end{array}
\eez
 \end{minipage} 
\begin{minipage}[b]{6cm}
\includegraphics[scale=.1]{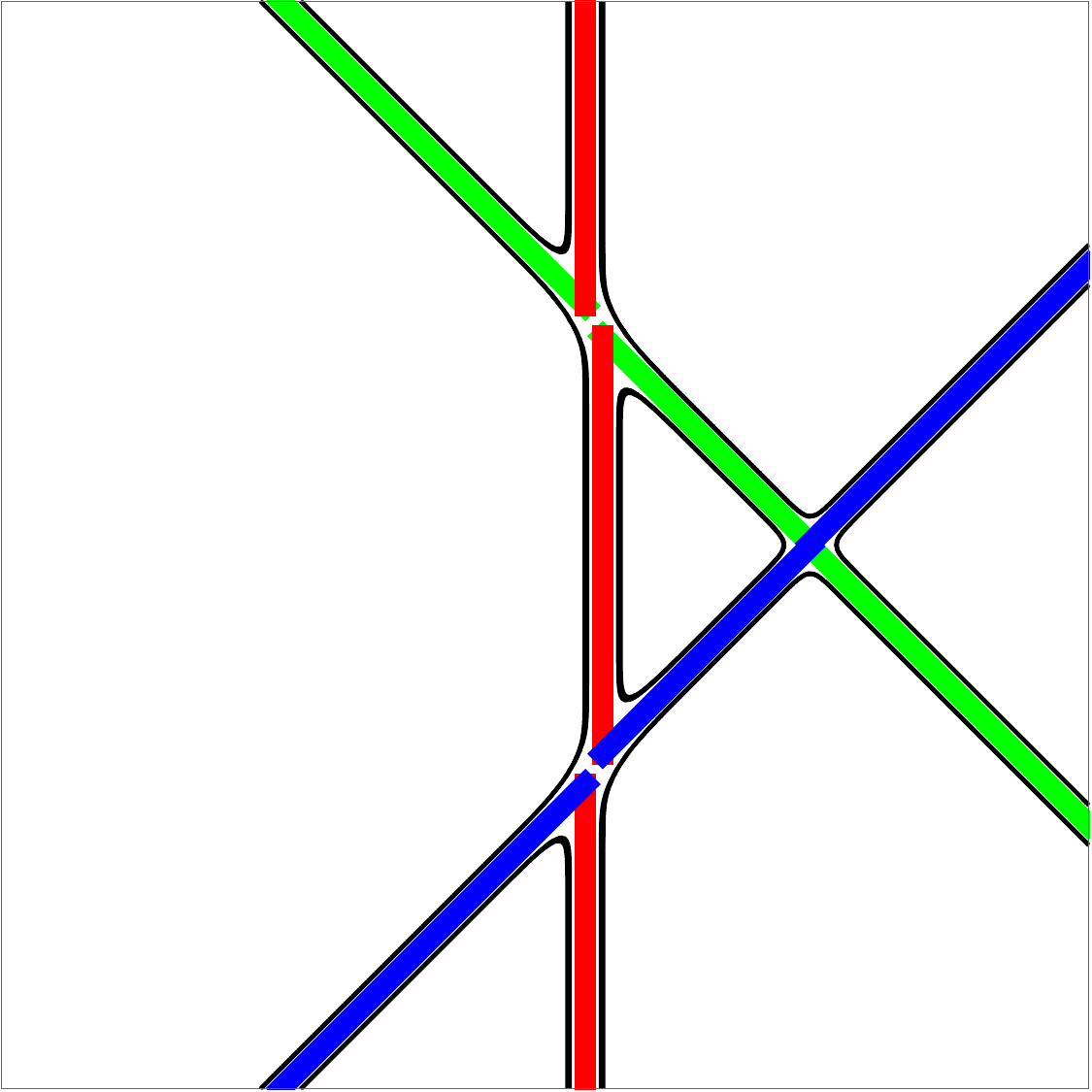}
\hspace{1cm}
\includegraphics[scale=.1]{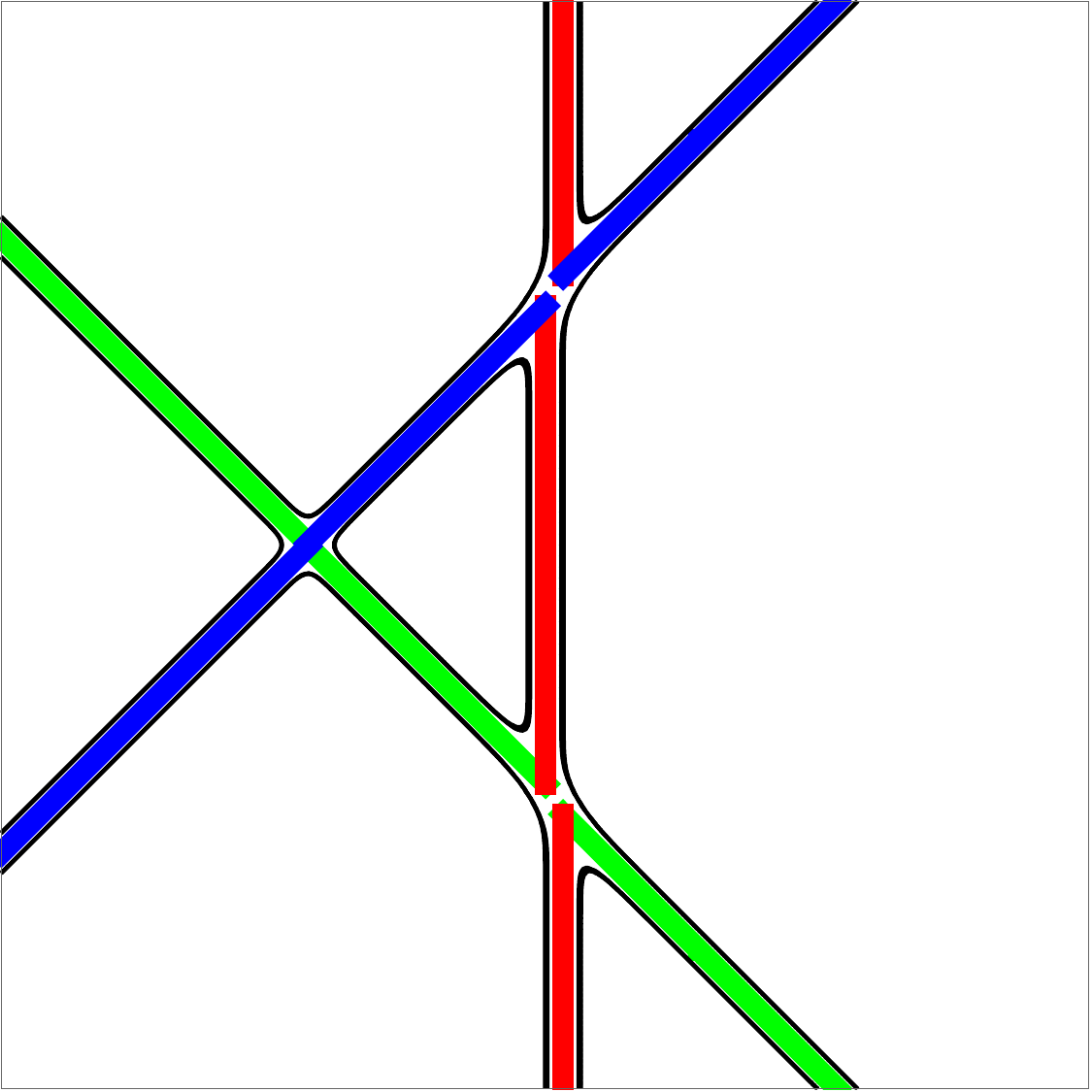}
\end{minipage} 
\caption{Contour plots of a 3-soliton solution of a matrix KP equation in the $x,y$-plane, at a negative and a positive value of time. 
These line soliton configurations realize left and right hand side of a familiar graphical representation of the Yang-Baxter equation. (The parameters of the solution have been tuned such that the soliton lines are arranged in this optimal way.) The fact that these two configurations are deformed into each other by the time evolution, implies that the two associated 
compositions of three 2-soliton data maps, acting at the intersections (of soliton lines) ordered in vertical ($y$) direction, take initial matrix data to equal data, since the maps do not depend on the coordinates $x,y,t$. This means that the Yang-Baxter equation holds. \label{fig:YB} }
\end{center}
\end{figure}

Bidifferential calculus does not only treat partial differential and difference equations on an equal footing, it is also suitable to deal with versions of integrable models, where the dependent variables take values in a noncommutative associative algebra. This includes matrix versions of soliton equations. Two asymptotically free matrix solitons will typically change their matrix data due to an interaction caused by the nonlinearity of the governing dynamical system. The 2-soliton interaction leads to a map $\mathcal{R}: \, \mathbf{S} \times \mathbf{S} \rightarrow \mathbf{S} \times \mathbf{S}$, where $\mathbf{S}$ is the set of allowed matrix data of a single soliton. It was known (for KdV and NLS) that such a map can be a solution of the (set-theoretic) \emph{Yang-Baxter equation}, which makes a statement about a 3-soliton interaction. The tropical limit provides us with a tool to concretely realize the Yang-Baxter relation (see Fig.~\ref{fig:YB}). It turned out, however, that even in the KdV case there are different maps obtained from 2-soliton interactions, of which only one satisfies the Yang-Baxter equation. In general, a 3-soliton interaction is constrained by a mixed (``intertwining") version of the Yang-Baxter equation, where also a map enters which is not a Yang-Baxter map \cite{DMH20b}. More generally, in \cite{DMH18,DMH19,DMHC19,DMH20b} the interaction of solitons of the matrix KP equation has been explored.

This sketch of Aristophanes' research surely conveys its variety and often striking novelty of underlying ideas and insights.

\renewcommand{\refname}{Publications of Aristophanes Dimakis}

\end{document}